# Theory of Space Magnetic Sail Some Common Mistakes and Electrostatic MagSail*

**Alexander Bolonkin**
C&R, 1310 Avenue R, #F-6, Brooklyn, NY 11229, USA
T/F 718-339-4563, aBolonkin@juno.com, http://Bolonkin.narod.ru

## Abstract

The first reports on the "Space Magnetic Sail" concept appeared more 30 years ago. During the period since some hundreds of research and scientific works have been published, including hundreds of research report by professors at major famous universities. The author herein shows that all these works related to Space Magnetic Sail concept are technically incorrect because their authors did not take into consideration that solar wind impinging a MagSail magnetic field creates a particle magnetic field opposed to the MagSail field. In the incorrect works, the particle magnetic field is hundreds times stronger than a MagSail magnetic field. That means all the laborious and costly computations in revealed in such technology discussions are useless: the impractical findings on sail thrust (drag), time of flight within the Solar System and speed of interstellar trips are essentially worthless working data! The author reveals the correct equations for any estimated performance of a Magnetic Sail as well as a new type of Magnetic Sail (without a matter ring).

## Introduction

The idea of utilizing the magnetic field to aggregate matter in space, harnessing a drag from solar wind or receiving a thrust from an Earth- charged particle beam is old. The MagSail is a gigantic (more than 50 -100 km in radius) super-conductive ring located in outer space that produces a magnetic field and reflects the impinging solar wind, or a particle beamed from the Earth. Unfortunately, the currently used theory for computation of drag from solar wind or thrust from particle beam is complex. The magnetic field changes in widely diapason, every particle moves in its own trajectory and it is exquisitely difficult to accurately estimate a summary drag. Over the years, many space researchers have offered many equations for drag estimation that give remarkably different results. However, no known equations take into proper consideration the magnetic field of particles moved in a ring-shaped magnetic field. These particles create their own magnetic field that is OPPOSED to the MagSail's magnetic field. This magnetic field of charged particles can be stronger—by hundreds times—than a ring field. It can fully deactivate the MagSail magnetic field.

The simplest computation shows a profound mistake in all known works. The reader can find part of them in [1]-[38], (see also [39]-[40]).

Take the typical MagSail ring: radius of ring is $R = 50$ km, electric current $I = 10^4$ A. The intensity $H_1$ of magnetic field in center of ring is

$$H_1 = \frac{I}{2R} = \frac{10^4}{2 \cdot 5 \cdot 10^4} = 0.1 \quad \text{A/m}, \tag{1}$$

This intensity is approximately same of the ring as well as near it. We assume in our subsequent computation that $H_1 =$ constant.

Take the typical solar wind flows into ring at distance from Sun 1 AU (the Earth's orbit about its primary star) with average wind speed $V = 400$ km/s, and density $N = 10^7$ 1/m$^3$. The solar wind contains electrons and protons. Within the ring magnetic field they rotate under Lawrence force and produce their own magnetic field that is OPPOSED to the ring magnetic field, decreases it (diamagnetic property), and pumps the ring magnetic energy into energy of its own magnetic field (summary energy is constant). This magnetic field from the rotated electrons (we here neglect the additional magnetic field from the rotated protons) can be estimated by equations (we consider only electrons into the ring):

$$H_2 = \frac{i}{2r}, \quad r = \frac{V}{(q/m_e)B_1}, \quad i = \pi R^2 qNV, \quad B_1 = \mu_0 H_1 \tag{2}$$

where $H_2$ is magnetic intensity from rotated solar wind electrons, A/m; $r$ is electron gyro-radius, m; $i$ is electric currency of solar wind electrons, A; $V = 400$ km/s is average solar wind speed, $B_1$ is magnetic intensity, T; $\mu_0 = 4\pi 10^{-7}$ is magnetic constant.

Substituting our values, we received $r = 18.2$ m; $i = 5024$ A; $H_2 = 276$ A/m. The last magnitude shows that the magnetic intensity of solar wind electrons is in 2760 times greater ($H_2 \gg H_1$) than the ring magnetic intensity of MagSail! It is correct for any charged beam that interacts with the MagSail. That means all research and computation (without an influence the solar wind or charged beam into MagSail) is wrong and basically worthless for all practical space exploration and exploitation applications.

How can it happen that hundreds of researchers, professors at famous universities, audiences of specialists, members of scientific Conferences and Congresses, editors of scientific journals: "Journal of Propulsion and Power" (Editor V. Yang); Journal "Spacecraft and Rockets", (Editor V. Zoby), paid so little attention to student-level mistakes in many scientific publications and public presentations to scientific conferences? More over, the director NASA Institute for Advanced Concepts (NIAC) Mr. R. Cassanova awarded (totaling more than $1 million dollars!) to his close associate, professor R.M. Winglee (University of Washington) for pseudo-scientific work about MagSail [1][1]. In eight years of NIAC's existence under him, Mr. Cassanova spent in excess of forty millions dollars of taxpayer money in pseudo-scientific works, but has not presented to the public even one new researched scientific concept. Since 90% of NIAC reports are pseudo-scientific or wrong works the Scientific Committee of a famous organization, the CAGW (Citizen Against Government Waste), awarded NIAC and Mr. Cassanova Pseudo-Nobel Prize-2005 [41]-[43].

It is still happening because popular textbook authors continue to consider the interaction between the strong magnetic field of particle accelerators and small amount of charged particles where we can neglect the influence of charged particles in magnetic field of the accelerator. With MagSail's, we have the opposed situation: the weak ring magnetic field and strong magnetic field of solar sail or charged beam.

## Theory

Below, the author suggests the correct theory of MagSail operation, which takes into consideration the influence of the solar wind flow into the ring magnetic field and allows an estimation of the drag of MagSail.

Let us to take the equations (2) in form:

$$H_1 = \frac{I}{R_1}, \quad H_2 = \frac{i}{2r}, \quad r = \frac{V}{(q/m_e)B},$$
$$R_2 = \frac{V}{(q/m_p)B}, \quad i = \pi R_3^2 qNV, \tag{3}$$
$$B = \mu_0(H_1 - H_2)$$

where $m_p$ is mass of positive particle, for proton $m_p = 1.67 \times 10^{-27}$, kg; $R_2$ is rotate radius of positive particles (protons for Solar Wind), m; $R_3$ is capture radius of positive particles, m.

Notice particularly the last equation (3). In this equation, the active is summary magnetic intensity $B$! For getting the maximum solar wind drag the turn radius of heavy particles must be 90 degrees. Assume $R=R_1=R_2=R_3$. We have 6 equations (3) and 6 unknown values. From set equations (3) we receive the estimation of the radius efficiency $R$:

$$R^2 = \frac{m_e}{\pi q^2 N}\left(\frac{Iq}{m_p V} - \frac{2}{\mu_0}\right), \tag{4}$$

From (4) we get minimal ring electric currency

$$I \geq \frac{2m_p V}{\mu_0 q}, \tag{5}$$

For average solar wind speed $V = 400$ km/s the minimal ring electric currency is $I = 6.65 \times 10^3$ A.
The solar wing drag, $D$, equals approximately

$$D = \pi R^2 m_p N V^2. \tag{6}$$

Results of computation are presented in figures 1 - 2. Look you attention: for receiving good drag we need in high electric current. For typical current $I = 10^4$ A ($I = 10$ kA) the efficiency radius $R$ and drag $D$ are small.

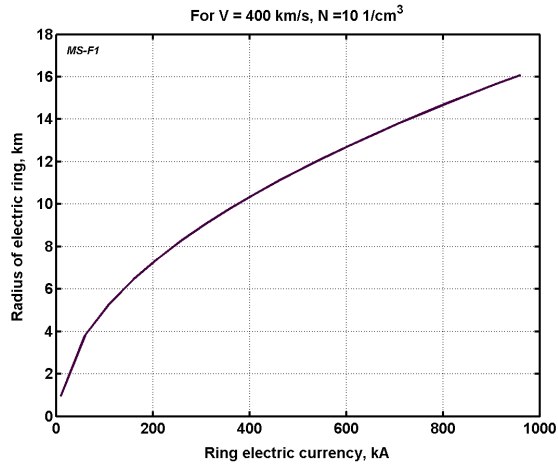

**Fig.1.** Radius efficiency of MagSail via ring electric current

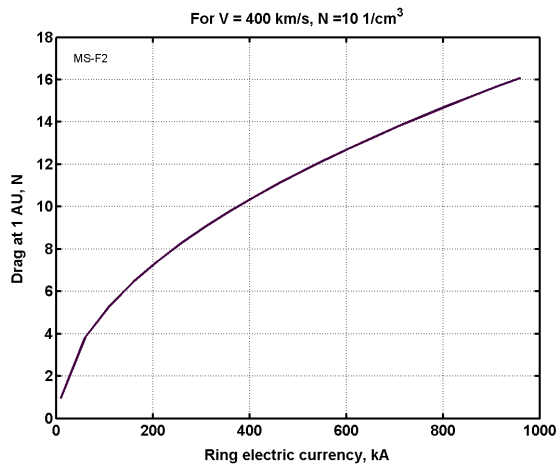

**Fig. 2.** Drag of MagSail via ring electric current at distance from Sun equals 1 Astronomical Unit.

## New Electrostatic MagSail (EMS)

The conventional MagSail with super-conductive ring has big drawbacks:
1. It is very difficult to locate gigantic (tens of km radius) ring in outer space.

2. It is difficult to insert a big energy into superconductive ring.
3. Super-conductive ring needs a low temperature to function at all. The Sun heats all bodies in the Solar System to a temperature higher then temperature of super-conductive materials.
4. The super-conductive ring explodes if temperature is decreased over critical value.
5. It is difficult to control the value of MagSail thrust and the thrust direction.

The author offers new Electrostatic MagSail (EMS). The innovation includes the central positive charged small ball and a negative electronic equal density ring rotated around the ball (fig.3).

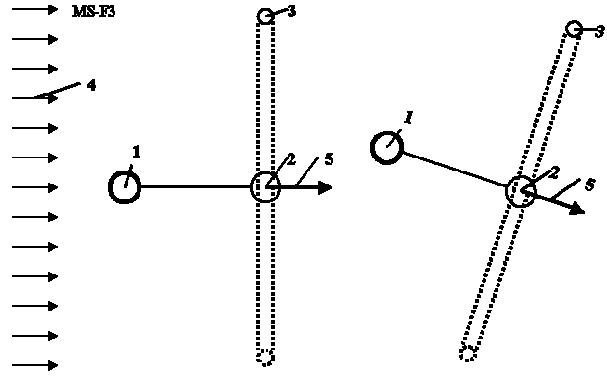

**Fig.3.** Electrostatic MagSail. Notations: 1 - Spaceship; 2 - Positive charged ball; 3 - electrical ring; 4 - solar wind; 5 - EMS drag. In right side the EMS in turn position.

The suggested EMS has the following significant advantages in comparison with conventional MagSail:
1) No heavy super-conductive large ring.
2) No cooling system for ring is required.
3) Electronic ring is safe.
4) The thrust (ring radius) easy changes by changing of ball charge.

## Electrostatic MagSail Theory

Let us consider a method of estimation of electronic ring magnetic intensity in the electronic ring's center [2]. We will take into consideration a repulsion of electrons from electron ring (blocking the ball charge by the electronic ring) and relativistic speed of electrons. We will not take into consideration diamagnetic property of solar wind or charged beam because our purpose here is only to find the magnetic intensity from electronic ring. The blocking the MagSail magnetic field by the particles flow the reader find in previous section (above). We also neglect the radiation of rotary electronic ring because the ring is right circle, has constant density and that does not emit synchronous radiation (this assumption needs further research. Synchronous radiation appears when electrons rotate in outer magnetic field, the electron ring is unclosed or has non-constant density. In our case the ring electric and magnetic fields are constant and not emit energy in outer space).
From equilibrium of the centrifugal and attraction forces we have

$$\frac{MV_e^2}{R} = k\frac{(Q_1 - Q_2)Q_2}{R^2},$$
$$M = m_e\frac{Q_2}{q}, \quad Q_1 > Q_2,$$

(7)

where $M$ is mass of electron ring, kg; $V_e$ is speed of electrons, m/s; $R$ is ring radius, m; $k = 9 \times 10^9$ is electrostatic constant; $Q_1$ is positive charge of the central ball, C; $Q_2$ is negative charge of the electron ring, C; $m_e$ is mass of electron, kg; $q = 1.6 \times 10^{-19}$ is electron charge, C.

The best relation between $Q_1$ and $Q_2$ is $Q_1 = 2Q_2$. Substitute this value into (7) we receive

$$V_e^2 = k\left(\frac{q}{m_e}\right)\frac{Q_2}{R}, \quad Q_2 = \frac{RV_e^2}{k(q/m_e)}, \quad H = \frac{I}{2R}, \quad I = \frac{Q_2 V_e}{2\pi R}, \quad B = \mu_0 H \tag{8}$$

where $I$ is ring electric currency, A; $H$ is magnetic intensity, A/m; $B$ is magnetic intensity, T; $\mu_0 = 4\pi 10^{-7}$ is magnetic constant.

Substitute the previous Eqs. (8) in the last equation (8) for $B$ and use the formula for relativistic electron mass

$$B = \frac{\mu_0}{4\pi R} \frac{V_e^3}{k(q/m_e)}, \quad \beta = \frac{V_e}{c}, \quad m_e = \frac{m_{e0}}{\sqrt{1-\beta^2}}, \quad B = \frac{\mu_0 c^3 (m_{e0}/q)}{4\pi R k} \frac{\beta^3}{\sqrt{1-\beta^2}} \tag{9}$$

where $c = 3 \times 10^8$ m/s is light speed; $m_{e0} = 9.11 \times 10^{-31}$ kg is electron mass at $V_e = 0$.

Let us to add formula for estimation charge and radius of ball and substitute the known values into last equation (9). We received the final equations for estimation of MagSail size:

$$B = 1.7 \cdot 10^{-3} \frac{1}{R} \frac{\beta^3}{\sqrt{1-\beta^2}},$$

$$H = \frac{B}{\mu_0} = 1.36 \cdot 10^3 \frac{1}{R} \frac{\beta^3}{\sqrt{1-\beta^2}}, \tag{10}$$

$$Q_2 = \frac{c^2 R}{k(q/m_{e0})} \frac{\beta^2}{\sqrt{1-\beta^2}}, \quad a^2 = \frac{2kQ_2}{E_0}$$

where $a$ is radius of ball, m; $E_0$ is safety electric intensity at ball surface, V/m.

If the magnetic intensity into ring is constant, we can estimate the energy needed for starting of ring:

$$H = \frac{I}{2R}, \quad \Phi = \mu_0 \frac{I}{2R} S = IL,$$

$$L = \mu_0 \frac{S}{2R} = \mu_0 \frac{\pi R}{2}, \quad W = \frac{LI^2}{2} \tag{11}$$

where $\Phi$ is magnetic flux, Wb: $L$ is ring inductance, Henry; $S$ is ring area, m$^2$; final equation in (11) $W$ is energy, J. For conventional ring of MagSail having $R = 50$ km and $I = 10^4$ A the $W = 5 \times 10^6$ J.

The Eqs. (7) - (11) allow to find magnetic intensity of MagSail for given ring radius and electron speed (without solar wind or plasma beam), charge and radius of ball for given electrostatic ball intensity, energy of rotate ring, but they do not permit to estimate a MagSail drag. We can estimate drag of conventional MagSail (see section above), to compute the drag of electrostatic sail offered by author in [3] Chapter 18, but unfortunately we cannot to estimate for the drag EMS in the time present. The trajectory of charged particle into both field (magnetic and electric) are very complex.

## Acknowledgement
The author wishes to acknowledge R.B. Cathcart for helping to correct the author's English.